\documentclass[aps,twocolumn,prd,showpacs,showkeys,preprintnumbers,superscriptaddress,nobibnotes,notitlepage,floatfix,longbibliography,nofootinbib,letterpaper,10pt]{revtex4-2}

\usepackage {amsmath, amssymb, graphicx, rotating, braket,mathtools}
\usepackage{mathrsfs}
\usepackage{comment}
\usepackage{mathtools}
\usepackage{aas_macros}
\usepackage[dvipsnames]{xcolor}
\usepackage[margin=1 in]{geometry}
\usepackage[colorlinks = true,
            linkcolor = Blue,
            urlcolor  = Green,
            citecolor = blue]{hyperref}

\def\be{\begin{equation}}
\def\ee{\end{equation}}
\def\bea{\begin{eqnarray}}
\def\eea{\end{eqnarray}}

\newcommand{\gsim}{\lower.7ex\hbox{$\;\stackrel{\textstyle>}{\sim}\;$}}
\newcommand{\lsim}{\lower.7ex\hbox{$\;\stackrel{\textstyle<}{\sim}\;$}}

\newcommand{\yr}{\rm yr}
\newcommand{\s}{\rm s}

\newcommand{\nhz}{\rm nHz}
\newcommand{\hz}{\rm Hz}

\begin{document}

\title{Constraining Attohertz Gravitational Waves \\ with the Integrated Sachs-Wolfe Effect}

\author{Jason Kumar}
\email{jkumar@hawaii.edu}
\affiliation{Department of Physics and Astronomy, University of Hawai'i, Honolulu, HI 96822, USA}

\author{Tao Xu}
\email{tao.xu@ust.hk}
\affiliation{Jockey Club Institute for Advanced Study, Hong Kong University of Science and Technology, Clear Water Bay, Kowloon, Hong Kong}
\affiliation{Department of Physics and Astronomy, University of Oklahoma, Norman, OK 73019, USA}

\begin{abstract}
We consider the scenario in which very low frequency (attohertz) gravitational waves (GWs) are produced 
in the epoch after recombination.  We show that such GWs can be constrained through their contribution to the integrated Sachs-Wolfe (ISW) effect on the cosmic microwave background. We derive constraints on the 
GW energy density which can be obtained from the ISW effect, without assuming a specific GW production mechanism. Our results demonstrate that the ISW effect can probe GW production in a frequency regime which cannot be accessed by conventional GW observatories.
\end{abstract}

\maketitle

\maketitle

\section{Introduction}

Over the past decade, revolutionary advances have been made in the detection of gravitational waves (GWs)~\cite{KAGRA:2021duu,NANOGrav:2023gor,EPTA:2023fyk,Antoniadis:2022pcn,Xu:2023wog}. 
Upcoming and planned GW observatories will increase current sensitivities by orders of magnitude~\cite{LISA:2024hlh,ET:2025xjr,Evans:2023euw,Sesana:2019vho,Corbin:2005ny,Weltman:2018zrl}.  
But these direct probes of 
GWs are generally restricted to operate in the frequency range of $\gtrsim \nhz$.  The reason is straightforward; if 
an observatory operates over a time scale of ${\cal O}(10^9 \s)$ ($\sim 30~\yr$), then it will only observe a full oscillation of a 
GW with frequency $\gtrsim 10^{-9}~\hz$.  But this argument assumes that the observation is based in the time domain.  
An alternative approach to probing GWs is based in the spatial domain, that is, observing the stretching and contraction of spacetime across 
astrophysical length scales.  Since observatories can probe length scales much larger than $30$ light years, this provides an 
opportunity to probe the effects of GWs with frequencies $\ll {\cal O}(\nhz)$.  In this work, we will consider 
a particular example of this general strategy: constraining the injection of GWs in the early universe through 
their distortions of the cosmic microwave background (CMB) via the integrated Sachs-Wolfe (ISW) effect~\cite{1967ApJ...147...73S}. 

The CMB provides a powerful channel for precision tests of cosmic evolution~\cite{Planck:2018vyg}. 
Observations of the full‑sky map, combined with high resolution on small angular scales, offer a unique probe of deviations from a perfectly homogeneous universe, particularly fluctuations in the late universe.
The ISW effect measures the distortion of the CMB caused by the redshift and blueshift of photons as they pass through 
time-varying gravitational potentials, such as those caused by GWs.  We will see that correlations between the ISW effect 
along different lines of sight can probe GWs with wavelengths much longer than $30$ light years, allowing us 
to probe GWs in the frequency range $10^{-19} - 10^{-17} \hz$.

The ISW effect has already been used to probe dark energy and the growth of structure 
(see, for example,~\cite{Granett:2008ju,Nadathur:2016hky,Naidoo:2022rda,Seraille:2024beb}), which 
generates time‑varying gravitational potentials 
that act as scalar‑type perturbations on the temperature of CMB photons.
GWs instead act as tensor‑type perturbations on the propagation of CMB photons~\cite{Nishizawa:2014vga}. 
A well‑known source of such metric fluctuations is cosmic inflation, where tensor fluctuations are generated, freeze out after horizon exit, and re‑enter the horizon at later times. Inflationary GWs have been probed through the B‑mode polarization of the CMB~\cite{PhysRevD.105.083524,Kusaka:2018yzq,PhysRevLett.127.151301,POLARBEAR:2022dxa,SPIDER:2021ncy,PhysRevD.101.122003,SPT-3G:2025vtb}, although a detection has yet to be made, and their impact on CMB temperature anisotropies remains negligible in standard inflationary models.

In principle, all tensor perturbations sourced during cosmic history leave imprints on the CMB sky. Going beyond inflationary GWs, there have been efforts to use precise cosmological observations to place constraints on tensor perturbations across a wide frequency range~\cite{Lasky:2015lej}. 
In particular, CMB observables are well suited to probing tensor perturbations on larger scales than current GW searches, so they provide a new channel for ultra-low-frequency GWs. Existing bounds on GWs from CMB B-mode measurements \cite{BICEP2:2018kqh, Clarke:2020bil,Ng:2021waj}, CMB spectral distortions \cite{Kite:2020uix}, and lensing of the 21-cm signal \cite{Book:2011dz} already illustrate this complementarity, although the current sensitivity has not yet reached meaningful regions of the model parameter space.  

Although previous studies have primarily focused on GWs generated in the early universe, particularly well before recombination, tensor perturbations produced at later times can also leave observable imprints on CMB photons as they propagate after their release at recombination. Here, we consider GWs generated after recombination. One possible observational signature of such late-time tensor perturbations is the CMB B-mode polarization induced during reionization. In particular, a B-mode power spectrum peaking at $\ell \lesssim 10$, without a corresponding peak at $\ell \sim 100$, could serve as a distinctive signature of late-time GW production~\cite{Geller:2021obo}. CMB B-mode observations can also probe tensor perturbations generated at epochs closer to recombination \cite{Greene:2024xgq,Ireland:2025yqr,Greene:2026one, Zebrowski:2026pye}.

We will focus on GWs which are produced after recombination, and their impact on the 
ISW effect~\cite{Bottaro:2026nna}.  An example of a process which could inject GWs after 
recombination is a cosmological first-order phase transition (FOPT) in the dark sector which percolates at relatively low temperature.  Although high-temperature FOPTs 
can produce GWs of high enough frequency to be detected at GW observatories, and FOPTs which occur during dark energy domination may be detected through 
their effect on cosmological evolution, FOPTs which occur between recombination 
and dark energy domination are more difficult to constrain~\cite{Koren:2025ymq}.  
But we will not focus on any particular scenario for the production of low-frequency GWs.  
Instead, we will assume the production of 
monochromatic GWs at some time after recombination, and determine the GW density fraction which would be allowed by current observational 
limits on the GW contribution to the ISW effect.  
For any particular scenario for the production of GWs with a particular spectrum, these results can 
then be convolved against the power spectrum and recast as a bound on the model. 

The plan of this paper is as follows.  In Section~\ref{sec:ISW} we discuss the contribution to the ISW 
effect of GWs produced after recombination.  In Section~\ref{sec:Results}, we present our 
constraints on low-frequency gravitational waves obtained from the ISW effect.  We conclude in 
Section~\ref{sec:Conclusion}.

\section{GW contribution to the ISW eﬀect}
\label{sec:ISW}

In this section, we review the mechanism by which the ISW effect is produced by GWs between recombination and the current epoch.

\subsection{Tensor-induced temperature anisotropy}

A Friedmann–Robertson–Walker (FRW) metric perturbed by a tensor mode may be written as  
\begin{equation}
    \mathrm{d}s^2 = a^2(\eta) \left[ -\mathrm{d}\eta^2 + \left( \delta_{ij} + h_{ij} \right) \mathrm{d}x^i \mathrm{d}x^j \right]
\end{equation}
where $\eta$ is the conformal time, and $h_{ij}$ represents the transverse and traceless tensor perturbation satisfying $\partial_i h_{ij}=0$, $h_{ii}=0$.
For a photon moving in the $\hat{p}^{i}$ direction in this background, the 
4-momentum can be written as $p^{\mu} = E(1,\hat{p}^{i})/a(\eta)$.
From the geodesic equation, at linear order in tensor perturbations,  the photon energy evolves as
\begin{equation}
    \frac{\mathrm{d} \ln(aE)}{\mathrm{d} \eta}
= -\frac{1}{2} h'_{ij} \, \hat{p}^i \hat{p}^j ,
\end{equation}
where 
$' = \partial / \partial \eta$
at fixed comoving position.
For CMB photons, the tensor perturbation introduces an additional redshift and blueshift of the photon energy along the line of sight, in analogy with the Sachs–Wolfe effect caused by a time-dependent gravitational potential (scalar mode contribution). Thus, tensor perturbations are expected to induce similar temperature anisotropies in the CMB sky map.

For the ISW effect to be non-trivial, it is necessary for the metric fluctuations to have a non-trivial time dependence.  
Otherwise, photons which blueshift as they fall into a gravitational potential well would redshift as they emerged 
from it, yielding no net ISW effect.  For this reason, linear matter perturbations do not contribute to the ISW 
effect during the matter-dominated epoch, because the associated gravitational potential remains constant.  
However, the metric perturbations associated 
with GWs will damp with the expansion of the universe, yielding a non-trivial ISW effect even during 
the matter-dominated epoch.

The CMB temperature anisotropy can be calculated with the line-of-sight integral in direction $\hat{\textbf{n}}$ from the recombination time $\eta_*$ (assuming instantaneous decoupling during recombination) to today $\eta_0$,
\begin{eqnarray}
\Theta^{(t)}(\hat{\mathbf{n}}) &\equiv& \frac{\delta T^{(t)}(\hat{n})}{\bar{T}} 
\nonumber\\ 
&=&  -\frac{1}{2} \int_{\eta_*}^{\eta_0} \mathrm{d}\eta \;
h'_{ij}(\eta, x(\eta)) \, \hat{n}^i \hat{n}^j
\nonumber\\
&=&-\frac{1}{2} \int_{\eta_*}^{\eta_0} \mathrm{d}\eta
\int \frac{\mathrm{d}^3 k}{(2\pi)^3} \,
h'_{ij}(\eta, \mathbf{k}) \,
\hat{n}^i \hat{n}^j \,
e^{i  k (\eta_0 - \eta) \hat{\mathbf{k}} \cdot \hat{\mathbf{n}}} ,\nonumber\\
\label{eq:ThetaGeneral}
\end{eqnarray}
where $h_{ij}(\eta, \mathbf{k})$ is the Fourier transform of the tensor perturbation, and we have 
assumed flat space.

\subsection{Tensor perturbation production and evolution}

We consider the case in which the GW is monochromatic in frequency and is injected in the post-recombination epoch 
at conformal time $\eta_i$. 
If the GW wavevector $\vec{k}$ is along the $z$-direction, then the tensor perturbation can then be expressed as
\begin{equation}
    h_{ij} =
\begin{pmatrix}
h_{+} & h_{\times} & 0 \\
h_{\times} & -h_{+} & 0 \\
0 & 0 & 0
\end{pmatrix} .
\end{equation}

As a simplification, we will assume that the universe is matter-dominated throughout the epoch relevant for the ISW effect.
Since we are not interested in the details of how the GWs are produced, we can simply assume that a 
source is turned on over a short period of time and generates GWs, which continue to propagate once the 
source is gone.  
An exact solution to the equation of motion, in a matter-dominated epoch and in the absence of a source, is 
\begin{eqnarray}
h_{\vec{k},\lambda} (\eta, \vec{x}) &=& h^0_{\vec{k},\lambda} 
\left[\frac{\sin{(k\eta + \alpha)}}{(k\eta)^3} - \frac{\cos(k\eta + \alpha)}{(k\eta)^2} \right] e^{\imath \vec{k} \cdot \vec{x} } 
\lambda_{ij},\nonumber\\
\label{eq:GWstrainsolution}
\end{eqnarray}
where  $\vec{x} = (\Delta \eta)  \hat n$, $\Delta \eta = \eta_0 - \eta$, and 
$\eta_0$ is the conformal time today.
$\lambda_{ij}$ is the polarization tensor (either $+$ or $\times$), and $\alpha$ is a constant which determines the phase of the solution.   We will take the polarization tensor to satisfy the normalization 
$\sum_{+,\times} \lambda^{ij} \lambda_{ij}^* = 4$.
The solution which is non-singular at early conformal times corresponds to $\alpha = 0$, but since the GWs were injected at 
finite time, $\alpha$ can assume any value between $0$ and $2\pi$.   However, the distinction between the 
solutions which are singular and non-singular at early time will be important if $k\eta \ll 1$.  
In this case, the regular solution is frozen out, while the singular solution damps rapidly.
Note, that after entering the horizon, the amplitudes of both modes scale as $\propto \eta^{-2} \propto a^{-1}$.

\subsection{ISW contribution to the CMB temperature anisotropy }

With the tensor perturbation evolution introduced above, the temperature anisotropy can be expressed as
\begin{widetext}
\begin{eqnarray}
\Theta^{(t)} (\hat n) &=& -\frac{1}{2}  
\int \frac{d^3 k}{(2\pi)^3} h^0_{(\vec{k},\lambda)} \lambda_{ij}\hat n^i \hat n^j
F(k, \hat k, \hat n, \eta_i, \alpha) ,
\eea
where
\bea
F(k,\hat k \cdot \hat n, \eta_i, \alpha) 
&=& e^{\imath k \eta_0  \hat k \cdot \hat n} \int_{k\eta_{i}}^{k\eta_0} dx 
\frac{\partial}{\partial x} \left[\frac{\sin (x+\alpha)}{x^3} - \frac{\cos (x+\alpha)}{x^2} \right] 
e^{- \imath x  \hat k \cdot \hat n} .
\end{eqnarray}
\end{widetext}
The temperature anisotropy is dominated by GWs propagating
transversely to the line of sight.

Expanding the temperature anisotropy in spherical harmonics 
following~\cite{Nishizawa:2014vga}
\begin{eqnarray}
\Theta_{\ell m} = \int d\Omega_{\hat n}~ \Theta (\hat n) Y_{\ell m} ( \hat n ),
\label{eq:Theta}
\end{eqnarray}
we obtain the angular power spectrum 
\begin{eqnarray}
C_\ell^\mathrm{ISW} = \langle \Theta_{\ell m} \Theta^*_{\ell m} \rangle .
\end{eqnarray}
We will assume a monochromatic GW power spectrum given by  
\begin{eqnarray}
\langle (h^0_{(\vec{k},\lambda)}) (h^0_{(\vec{k}',\lambda')})^*  \rangle &=& 
(2\pi)^3 P_h (k) \delta^3 (\vec{k}-\vec{k}') \delta_{\lambda \lambda'} ,
\nonumber\\
P_h (k) &=& \Delta^2 \delta (k-k_0) ,
\end{eqnarray}
where $k_0$ is the comoving wavenumber of the monochromatic GW signal, and 
$\Delta^2$ is a normalization which sets the GW energy density ($\Delta$ has units of 
$\mathrm{length}$).  

We will assume that for each GW direction $\hat k_0$, the associated phase $\alpha$ is essentially drawn uniformly from $[0,2\pi)$.\footnote{This phase does 
not affect the amplitude of the wave after it enters the horizon.}  
We should then average over $\alpha$ when computing $C_\ell^\mathrm{ISW}$. 
We then have
\begin{widetext}
\begin{eqnarray}
C_\ell^\mathrm{ISW} 
&=&  \frac{\Delta^2 k_0^2}{2} 
\left( \frac{1}{4\pi} \frac{(\ell - 2)!}{(\ell + 2)!} \right) \left[ \frac{1}{2\pi} \int_0^{2\pi} d\alpha  \left|\int d\cos \theta~  \sin^2 \theta 
P_\ell^{m=2} (\cos \theta) 
F(k_0,\cos \theta, \eta_i , \alpha) \right|^2 \right] .
\label{eq:ClGW}
\end{eqnarray}
\end{widetext}
Note that, in the angular integrals over $\hat n$ in Eq.~\eqref{eq:Theta}, the only dependence of 
$\Theta (\hat n)$ on the azimuthal angle arises from the polarization tensor contraction 
$\lambda_{ij} \hat n^i \hat n^j$, which picks out the $m=\pm 2$ spherical harmonics.  
In particular, we have a vanishing contribution for $\ell < 2$.

Our expression is a nested integral over the angular scale $\theta$ and the 
conformal time $\eta$.  It will be useful to perform the angular integral first, yielding 
\begin{widetext}
\bea
C_\ell^\mathrm{ISW} (\Delta, k_0) &=&  \frac{\Delta^2 k_0^2}{2} 
\left( \frac{1}{4\pi} \frac{(\ell - 2)!}{(\ell + 2)!} \right)
\left[ \frac{1}{2\pi} \int_0^{2\pi} d\alpha  \left|
\int_{k_0 \eta_{i}}^{k_0 \eta_0} dx ~ I(x,\alpha) 
\times G_\ell (k_0 \eta_0 - x)  \right|^2 \right] ,
\label{eq:ClGW_2}
\eea
where
\bea
I(x,\alpha) &=& \frac{\sin (x+ \alpha) }{x^2}  + 3 \frac{\cos (x + \alpha)}{x^3}  -3 \frac{\sin (x + \alpha)}{x^4}  ,
\eea
\end{widetext}
encodes the dependence of the gravitational potential on conformal time, and 
\bea
G_\ell(y) &=& \int_{-1}^1 dz~ (1-z^2) ~P_\ell^{m=2} (z) ~ e^{\imath y z} ,
\eea
is an integral which can be performed analytically, and encodes the contribution of anisotropies 
with comoving length scale $k_0^{-1}$ at conformal time $\eta$ on 
the angular scale parameterized by $\ell$.  
Note that this simple and computationally tractable form of $C_\ell^\mathrm{ISW}$ arises because we have 
assumed a matter-dominated universe.

Although the exact expression for $G_\ell$ is not particularly 
enlightening, we find for large $y$ and small $\ell$ that $G_\ell (y) \propto y^{-3}$. 
Taking $y = k_0 (\eta_0 - \eta) \sim \theta^{-1}$, where $\theta$ is the angular scale subtended 
by a transverse mode,
we see that the limit $y \gg 1$ corresponds 
to early conformal times, for which the angular scale subtended by transverse waves of comoving 
wave number $k_0$ is small, yielding a suppressed contribution to anisotropies at small $\ell$. 

Taking 
$x = k_0 \eta$ and $y = k_0 \Delta \eta = k_0 (\eta_0 - \eta)$, we see that the ISW effect is 
suppressed at times well after horizon crossing ($x \gg 1$) by the damping of the GW amplitude, 
and is suppressed at times well before the current epoch ($y \gg 1$) by the angular integral.  
When $k_0 \eta_i  \gg 1$, the angular suppression will dominate, and the dominant contribution 
to ISW effect will come from late times, when modes of wavenumber $k_0$ subtend a large angular scale.
In this limit, we find $y \ll 1$ and $x \sim k_0 \eta_0$, implying that  
$C_\ell^\mathrm{ISW} \propto (k_0 \Delta)^2 (k_0 \eta_0)^{-4}$.

But if the injected modes are already outside the horizon ($k_0 \eta_i \ll 1$) the dominant contribution 
to the ISW effect occurs right after injection, and arises from the sharp damping of the 
solution which is not regular as $\eta \rightarrow 0$.  
However, we have ignored the process which injected the GWs, and we would expect the 
ISW effect arising from the initial injection of GWs (when the amplitude is increasing) to 
tend to cancel the ISW effect arising from the subsequent damping of the GW amplitude.  Since the dominant 
contribution to the ISW effect occurs near the time of injection when $k_0 \eta_i \ll 1$, and we 
do not choose any particular model for the injection process (and thus cannot model its contribution 
to the ISW effect), we restrict ourselves to $k_0 \eta_i \geq 1$.

For a general choice of the GW power spectrum $P (k_0)$, we can obtain the ISW contribution 
to the temperature anisotropy power spectrum by convolving Eq.~\eqref{eq:ClGW_2} with the 
power spectrum, yielding
\bea
C_\ell^\mathrm{ISW} &=& \frac{1}{\Delta^2} \int dk_0~P (k_0) ~ C_\ell^\mathrm{ISW} (\Delta, k_0) . 
\label{eq:ClISW_general}
\eea
However, as we remain agnostic regarding the GW source, 
we will assume  a monochromatic GW spectrum.
We will be able to constrain $\Delta$ using the observational uncertainty in $C_\ell^\mathrm{ISW}$, and 
will recast the constraint on $\Delta$ into an upper bound on the GW energy density.

\subsection{Relation to the GW energy density}

We can find the energy density of a GW in the current epoch (when spacetime is close to 
Minkowski) using the Isaacson stress energy tensor
\begin{eqnarray}
T_{\mu \nu} &=& \frac{1}{32\pi G_N} \langle \partial_\mu h_{\alpha \beta} \partial_\nu h^{\alpha \beta} \rangle ,
\end{eqnarray}
where the average is taken over a spacetime region which is large compared to the wavelength/period.  In 
order for this quantity to be meaningful, the GW must enter the 
horizon before the current epoch.
Taking $k\eta_0 \gg 1$, where $\eta_0$ is the conformal time today, 
we find 
\begin{eqnarray}
\Omega_{\rm GW,0} &=& \frac{\Delta^2 }{12\pi^2 H_0^2 c^2}   
\frac{ 1}{\eta_0^4} .
\label{eq:OmegaGW0}
\end{eqnarray}
We thus see that if GWs of comoving wavenumber 
$k_0$ are injected at conformal time $\eta_i$, the GW energy density fraction in the current epoch must 
satisfy 
\begin{widetext}
\bea
\Omega_{\rm GW,0} &<& \frac{\delta C_\ell}{6\pi^2 (H_0 \eta_0)^2 (k_0 \eta_0 c)^2}  
\left( 4\pi \frac{(\ell + 2)!}{(\ell - 2)!} \right) 
\nonumber\\
&\,& \times 
\left[ \frac{1}{2\pi} \int_0^{2\pi} d\alpha  \left|
\int_{k_0 \eta_{i}}^{k_0 \eta_0} dx 
\left[\frac{\sin (x+ \alpha) }{x^2}  + 3 \frac{\cos (x + \alpha)}{x^3}  -3 \frac{\sin (x + \alpha)}{x^4} \right] 
\times G_\ell (k_0 \eta_0 - x)  \right|^2 \right]^{-1},
\label{eq:Bound}
\eea
\end{widetext}
where $\delta C_\ell$ is the bound on the contribution to the angular 
power spectrum from the ISW effect, and is given by the uncertainty in the observational constraints.
The GW energy density at injection and at later times is related to the energy density now just 
by the redshift factor $a^{-4}$.

\section{Results}
\label{sec:Results}

We consider the scenario in which GWs with a monochromatic spectrum are injected 
at conformal time $\eta_i$.  Most sources of GWs (for example, a cosmological 
first-order phase transition) will produce GWs with a continuous spectrum over 
a range of time, but our constraints can be recast for any particular model by convolving against the 
power spectrum, as in Eq.~\eqref{eq:ClISW_general}.

\begin{figure}[t]
\centering
\includegraphics[width=0.95 \columnwidth]{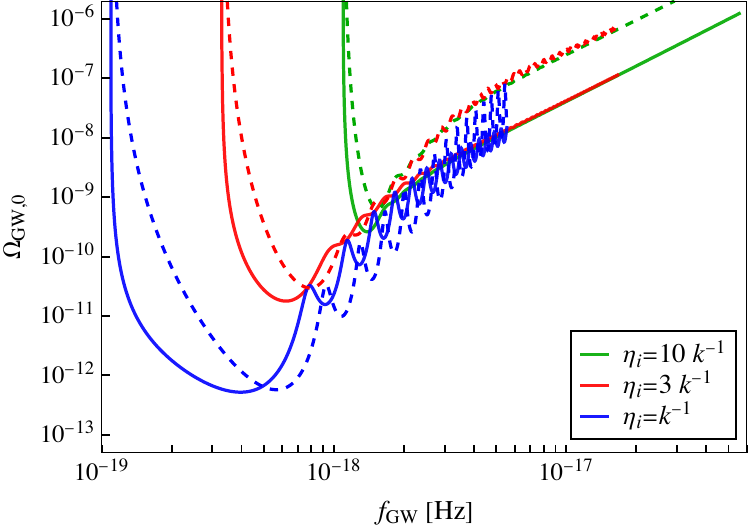}
\caption{Present-day monochromatic GW energy density, plotted as a function of GW frequency, required to produce a variation in temperature anisotropy given by
$D_{\ell=2}^\mathrm{ISW} =533.1~\mu\textrm{K}^2$ (solid curves) and 
$D_{\ell=3}^\mathrm{ISW} =1212.3~\mu\textrm{K}^2$ 
(dashed curves), with both values saturating the uncertainty levels of the Planck measurements at the corresponding multipoles \cite{Planck:2019nip}. 
The comoving wavenumber $k$ is related to the GW frequency via $f_{\textrm{GW}}=k/2\pi$.  
Each curve corresponds to a fixed value of $k \eta_i$, labeled by color.  
For each curve, a given GW frequency
thus corresponds to a different GW injection time relative to the horizon reentry time.
}
\label{fig:OmegaGWSensitivity}
\end{figure}

In Figure~\ref{fig:OmegaGWSensitivity}, we plot constraints on $\Omega_{\rm GW}$ obtained 
by requiring the temperature anisotropy $D_{\ell}\equiv [\ell(\ell+1)/2\pi] \, C_{\ell}$ be consistent 
with the Planck constraints  $D_{\ell=2}^\mathrm{ISW} < 533.1~{\rm \mu K}^2$ (solid) and  
$D_{\ell=3}^\mathrm{ISW} < 1212.3~{\rm \mu K}^2$ (dashed).   
These values represent the upward observational uncertainty in the angular 
power spectrum at $\ell=2, 3$~\cite{Planck:2018vyg}.  
We assume GWs of fixed comoving wavenumber $k$ are injected at conformal time $\eta_i$.  
Each contour represents a fixed choice of $k \eta_i$, as labeled.  The GW frequency today is given by 
$f_{\rm GW} = k/2\pi$.  We assume that GWs are injected between the conformal times of recombination and 
now ($\eta_* \leq \eta_i \leq \eta_0)$.  
But we require that $k \eta_0 \geq 1$, that is, that the wavelength is shorter than the Hubble scale.  
This constraint limits our analysis to the frequency range $f_{\rm GW} \gtrsim 10^{-19}~\hz$.
For each contour of fixed $k \eta_i \geq 1$, this constraint 
determines the range of accessible frequencies; the right edge of the contour corresponds to $\eta_i = \eta_*$, 
while the left edge of the contour corresponds to $\eta_i = \eta_0$.  So for each curve, from left to right, 
the frequency changes by a factor of $\eta_0 / \eta_* \sim 50$.

For contours with fixed $k \eta_i \gg 1$, the upper bound on $\Omega_{\rm GW}$ scales as $f_{\rm GW}^2$.
This result follows from the considerations described above: for modes which are within the 
horizon at the time of injection, the dominant contribution to the ISW effect occurs at late times, 
and $C_\ell^\mathrm{ISW} \propto (k \Delta)^2 / (k \eta_0)^{4}$.  We then find the bound
$\Omega_{\rm GW} \propto \Delta^2 \propto C_\ell^\mathrm{ISW} f_{\rm GW}^2$.  In particular, since 
the dominant contribution to the ISW effect occurs at late times, the bound is relatively 
insensitive to the injection time, and to the precise mechanism by which GWs are 
injected.
We see from Figure~\ref{fig:OmegaGWSensitivity} that models with $\Omega_{\rm GW} \sim 10^{-12}$ can 
be constrained through the ISW effect.
By comparison, CMB B-mode constraints on inflationary tensor modes are at the level of $\Omega_{\rm GW} \sim 10^{-10}$ to $10^{-14}$, 
over the range $f_{\rm GW} \sim 10^{-19}~\hz$ to $10^{-17}~\hz$~\cite{Lasky:2015lej}.

Note that, in the matter-dominated universe, $\Omega_{\rm GW}$ scales as $\propto a^{-1}$ for sub-horizon modes, implying that  
$\Omega_{\rm GW}$  at any point after recombination can be no more than a factor of $\sim 10^3$ larger than it is now.  
For points where $\Omega_{\rm GW,0} < 10^{-5}$, we know that GWs could never have constituted more than about 
$1\%$ of the energy density of the universe.   As a point of comparison, if the GWs were injected at recombination with an energy density equivalent to $\Delta N_{\rm eff} \sim 0.33$, then the energy 
density fraction in the current epoch would be  $\Omega_{\rm GW,0}\sim 10^{-5}$.

\section{Conclusion}
\label{sec:Conclusion}

We have considered a strategy for constraining  gravitational waves injected after recombination whose 
frequencies lie in the attohertz range.  This frequency range is far lower than the 
range which can be probed by GW observatories, whose sensitivity is limited by their exposure time 
of ${\cal O}(10)$ years.  Instead, we use the spatial domain, relying on effects integrated over long 
distance scales to probe very low frequency gravitational waves.  In particular, we utilize the 
ISW effect; long wavelength GWs will modify the red/blueshifting of CMB photons on large angular scales.

We consider the scenario in which gravitational waves are injected at some conformal time 
$\eta_i$ after recombination, when we take the universe to be matter-dominated.  
Our focus is on the impact of the ISW effect on temperature 
anisotropies at low $\ell$.  
The ISW effect is particularly interesting in this context because it is sensitive to GWs injected 
at any time after recombination.  By contrast, although B-mode polarization studies are 
very sensitive to GWs injected before recombination, they are of less utility for GWs injected 
between recombination and reionization.
If the injected GW is already within the horizon at the time 
it is injected ($k \eta_i \gg 1$), the dominant contribution to the ISW effect at small $\ell$ 
comes at late conformal times, and sensitivity is stronger at lower frequencies.

In order to make the computation analytically tractable, we have assumed a matter-dominated universe throughout 
the epoch after recombination.  
But for modes which are injected well inside the horizon, the largest contribution to the ISW effect occurs 
at late times, when dark energy is relevant.  A detailed numerical calculation of the associated corrections 
would be interesting, but would be computationally intensive, and is beyond the scope of this work.

We have assumed a monochromatic GW spectrum at injection, in order to present a sensitivity to gravitational 
waves which is comparable to that presented by GW observatories.  Generally, any particular GW source will 
produce a GW spectrum which depends in detail on the underlying physics.  Given a source, one could perform 
the appropriate integral over its spectrum to determine the contribution to  $C_\ell^\mathrm{ISW}$, allowing a 
more detailed analysis of any particular scenario.  It would be interesting to consider how well one can constrain 
particular sources, such as cosmological dark sector first order phase transitions, with the ISW effect.

We have only determined constraints using the lowest non-trivial multipoles of the temperature anisotropy.  
As we have seen, since higher-$\ell$ multipoles are sensitive to smaller angular scales, they may 
provide tighter constraints on higher frequency GWs. A global likelihood analysis 
utilizing the higher multipoles would be very useful, especially in conjunction with 
a full model for the GW spectrum, and with data from the next generation of CMB experiments~\cite{CMB-S4:2022ght}. We leave a detailed analysis of this approach and their potential sensitivities to future work.

Two sources of scalar perturbations may provide additional contributions to the ISW effect. GWs produced after recombination can source scalar perturbations through their energy density. This effect is expected to be less significant since it arises at second order. The underlying GW injection mechanism may also generate scalar perturbations, whose ISW contribution is subdominant during matter domination. These contributions could nevertheless be relevant in some scenarios, and quantifying them would be an interesting direction for future work.

Finally, we note that this work presents just one example of a general approach to probing sub-nanohertz gravitational 
waves, namely, using the spatial domain.  In particular, observing variations in the ISW effect over large angular 
scales provides sensitivity to transverse gravitational waves with very large wavelengths, thus accessing frequencies well 
below what could be directly observed over a timescale of a few decades.  It would be interesting to apply this general 
approach to other cosmological observables in order to find new windows into very low frequency gravitational waves.

{\bf Acknowledgments}  We are grateful to William Giare, Yuxin Liu, Shengyi Liu, Jeremy Sakstein, Kuver Sinha, and Yue Zhao for helpful discussions.  
JK and TX wish to acknowledge the Center for Theoretical Underground Physics and Related Areas (CETUP*), the Institute for Underground Science at Sanford Underground Research Facility (SURF), and the South Dakota Science and Technology Authority for hospitality and financial support, as well as for providing a stimulating environment.
JK is supported in part by DOE grant DE-SC0010504. TX acknowledges support from the HKUST Jockey Club Institute for Advanced Study.

\bibliographystyle{JHEP}
\bibliography{reference}

\end{document}